\documentclass[12pt]{article}
\usepackage{epsfig}
\usepackage{amssymb}
\usepackage{amsmath,amsfonts,amssymb,graphicx}
\newcommand{\beq}{\begin{eqnarray}}
\newcommand{\eeq}{\end{eqnarray}}

\begin{document}
\parindent 5mm
\vspace{-20ex} \vspace{1cm}
\begin{center}
\centerline{\Large \bf Quark Virtuality Distribution in
Non-perturbative }
\vspace{0.4cm}
\centerline {\Large \bf QCD Vacuum$^{*}$}

\footnote {$^{*}$ {\footnotesize Supported in part by National
Natural Science Foundation of China: 10647002, Guangxi Science
Foundation for Young Researchers:0991009, Guangxi Natural Science
Fuundation: 2011GXNSGA018140, Department of Guangxi Education for
the Excellent Scholars of Higher Education: 2011-54, Doctoral
Science Foundation of Guangxi University of Technology : 11Z16, and
also Pittsburgh Foundation, USA.}}
\end{center}

\vspace{0.4cm} \centerline{\large \bf Zhou Li-juan$^{1}$,~Ma
Wei-xing$^{2}$,~Leonard S. Kisslinger$^{3}$}

\vspace{0.6 cm} \centerline{$^1$ Collaboration group of hadron
physics and non-perturbative QCD study, Guangxi University
} \centerline{of Technology, Liu Zhou, 545006, Guangxi, China}
\vspace{0.4cm}
\centerline{$^2$ Institute of high energy physics, Chinese Academy
of Sciences, P. O. Box 918 Beijing,} \centerline{100049, China}
\vspace{0.4cm}
\centerline{$^{3}$ Department of Physics, Carnegie-Mellon
University, Pittsburgh, PA.15213, USA}

\vspace{0.6cm}
\begin{center}
{\Large\bf Abstract}
\end{center}
\vspace{0.6cm}

  We introduce the quark virtuality distribution [$\lambda^{2}_{q}(x)$]
as an extenson of the quark virtuality by using nonlocal rather than
local quark vacuum condensates. 
Our results show that the quark virtuality
distribution decreases gradually from a finite value to zero as $x$
increases. This clearly indicates that quarks are confined inside
hadrons, whereas, as $x$ approaches zero quarks have a finite
virtuality, which shows asymptotic freedom of quarks in QCD. The
present predictions are important not only for studying the
properties of QCD vacuum states, and non-perturbative QCD problems,
but also for study of the cosmological constant problem, which we have shown
should be treated using nonlocal quark vacuum condensate in the Ghost 
theory of QCD.

\vspace{1cm}
\vspace{0.4cm} {\bf Key Words}: {\footnotesize quark virtuality
distribution, non-perturbative QCD, fully dressed quark propagator.

\vspace{0.4cm} {\bf PACS} Number(s): 12.38.Aw,12.38.Lg,14.65.-q,98.80.-k}

\newpage
\section {Introduction }

  The quark virtuality, an expression for the momentum carried by
the quark vacuum condensate, was introduced during the early study of
QCD sum rules\cite{bi88,mr89,br91}, and was calculated\cite{s01,1} using the 
local quark condensate and local mixed quark-gluon condensate calculated using 
Dyson-Schwinger equations\cite{2a,2b}. Our present estimate of the quark 
virtuality distribution is based
on our previous calculations of quark virtuality and the nonlocal quark 
condendate used in our recent work on the cosmological constant\cite{3}.

Studying the quark virtuality distribution in non-perturbative QCD
vacuum state is of paramount importance for present day particle
and nuclear physics, since it is not only related to the property
of QCD vacuum states but also to studying non-perturbative QCD issues,
as well as the cosmological constant problem.

In Sect.2 we first review the quark local vacuum condensates and the
calculation of the quark virtuality in our previous work, and then 
calculate the quark virtuality distribution using a nonlocal quark 
condenensate. Finally, concluding remarks are given in Sect.3.

\section {The quark virtuality and quark virtuality distribution}

 In the first subsection we review the quark condensates virtuality, 
$\lambda_q^2$,
which was estimated in Ref.\cite{1}. We then derive the quark virtuality
distribution using a nonlocal quark condensate related to that used
in Ref.\cite{3}.

\subsection{The quark virtuality}

  The quark virtuality is defined in terms of the local quark condensate
and mixed quark-gluon condensate. These are given by the Taylor expansion
of the quark propagator, which we now review.
\begin{eqnarray}
S_q(x)=<0|:\hat{T}\bar{q}^{a}(x)q^{b}(0):|0>
\end{eqnarray}
where $q^{a}(x)$ is quark field with color index $a$, and $\hat{T}$
is the time-ordering operator. The fully dressed quark propagator in
Eq.(1) can be decomposed into a perturbative part and a
non-perturbative part, with only the non-perturbative part needed for the
present work. The  non-perturbative part of the quark propagator is
\begin{eqnarray}
\label{NP}
S^{NP}_q(x)=-\frac{1}{12}[<0|:\bar{q}(x)q(0):|0>+ 
x_{\mu}<0|:\bar{q}(x)\gamma^{\mu}q(0):|0>]
\end{eqnarray}
in configuration space, with a sum over color implied. $\gamma^{\mu}$ are
Dirac matricies. For short distances $x$, the Taylor expansion in $x$
of the scalar part of $S^{NP}_{q}(x)$, $<0|:\bar{q}(x)q(0):|0>$, is\cite{2a}
\begin{eqnarray}
\label{barqx}
<0|:\bar{q}(x)q(0):|0>=<0|:\bar{q}(0)q(0):|0>-\frac{x^{2}}{4}
<0|:\bar{q}(0)[ig\sigma G]q(0):|0> + \cdots
\end{eqnarray}
In Eq.(\ref{barqx}) the quantities in the expansion are the local quark
vacuum condensate, the quark-gluon mixed local vacuum condensate
and so forth. The $G$ in the second term of Eq.((\ref{barqx}) is the gluon 
field strength tensor, where the Dirac and color indexes have not been shown 
explicitly.

  In Ref.\cite{1} the Dyson-Schwinger Equations\cite{2a,2b} were used to
derive the local quark vacuum condensate $<0|:\bar{q}(0)q(0):|0>$ and the
local mixed quark-gluon condensate, $<0|:\bar{q}(0)[ig\sigma G]q(0):|0>$.
See  Ref.\cite{1} for these local condensates for three sets of
quark interaction parameters.

We review the definition of quarks mean
squared momentum in non-perturbative QCD vacuum, which is also
called the quark virtuality, $\lambda^{2}_{q}$. The
quark virtuality $\lambda^{2}_{q}$ is defined\cite{qprop,s04} as
\begin{eqnarray}
\label{lambda}
\lambda^{2}_{q}\equiv\frac{<0|:\bar{q}D^{2}q:|0>}{<0|:\bar{q}q:|0>}
\end{eqnarray}
where $D_{\mu}=\partial_{\mu}-ig_{s}A_{\mu}$, with
$A_{\mu}=A^{a}_{\mu}\lambda^{a}/2$, is the covariant derivative and
$\lambda^{a}$ is a $SU_{c}(3)$ Gell-Mann matrix. As shown in 
Refs.\cite{qprop,s04}, $<0|:\bar{q}D^{2}q:|0>$ can be written
\begin{eqnarray}
\label{d2q}
<0|:\bar{q}D^{2}q:|0>=-m^{2}_{q}<0|:\bar{q}(0)q(0):|0>+
\frac{1}{2}<0|:\bar{q}(0)[ig_{s}\sigma G]q(0):|0>.
\end{eqnarray}
Thus, from Eq.(\ref{d2q}) the quark virtuality $\lambda^{2}_{q}$ can be 
rewritten as
\begin{eqnarray}
\label{lq2}
\lambda^{2}_{q}=\frac{1}{2}\frac{<0|:\bar{q}(0)[ig_{s}\sigma
G]q(0):|0>}{<0|:\bar{q}(0)q(0):|0>}-m^{2}_{q}.
\end{eqnarray}
For light quarks $u$,$d$ and $s$, the mass $m^{2}_{q}$ term in Eq.(\ref{lq2}) 
is so small that it can be neglected. Therefore, we finally arrive at
\begin{eqnarray}
\lambda^{2}_{q}\simeq\frac{1}{2}\frac{<0|:\bar{q}(0)[ig_{s}\sigma
G]q(0):|0>}{<0|:\bar{q}(0)q(0):|0>}.
\end{eqnarray}

  The quark virtuality for three sets of parameters is given in Ref\cite{1}

\subsection {Quark virtuality distribution in the non-perturbative 
QCD vacuum }

 We define the quark virtuality distribution $\lambda^{2}_{q}(x)$ in the
non-perurbative QCD vacuum as a natural extension of Eq(\ref{lambda}) by
\begin{eqnarray}
\label{lambdax}
\lambda^{2}_{q}(x)\equiv\frac{<x|:\bar{q}D^{2}q:|0>}{<0|:\bar{q}q:|0>}
\; ,
\end{eqnarray}
similar to the extension of the quark condensate to the nonlocal
quark condensate in Refs.\cite{jk95,jk98}.
In Ref.\cite{jk95} $<x|:\bar{q}D^{2}q:|0>$ was evaluated, giving
\beq
\label{d2qx}
 <x|:\bar{q}D^{2}q:|0>&=& \frac{g_\lambda(x^2)}{2}
<0|:\bar{q}(0)[ig\sigma G]q(0):|0> \; ,
\eeq
with
\beq
\label{glx}
 g_\lambda(x^2)&=& \frac{1}{1+\lambda^2 x^2/8} \; .
\eeq

   The quantity $\lambda^2$ in Eq(\ref{glx}) is $\lambda^2 \simeq 0.3-0.8
{\rm GeV}^2$. From Eqs(\ref{lambdax}, \ref{d2qx}, \ref{glx}, \ref{lq2})
\beq
\label{lambda-x}
 \lambda^{2}_{q}(x) &\simeq& \frac{1}{1+\lambda^2 x^2/8}\lambda^{2}_{q}
\; .
\eeq

For the present work we take $\lambda^2 = 0.8 {\rm GeV}^2$, and
$ g_\lambda(x^2)$ is shown in the figure.

\clearpage

\begin{figure}[ht]
\begin{center}
\epsfig{file=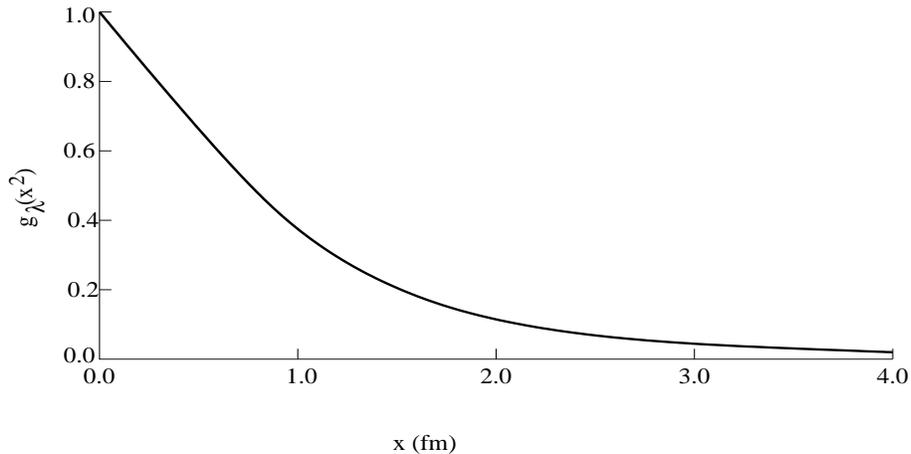,height=6cm,width=12cm}
\end{center}
\caption{$x$-dependence of the function $g_\lambda(x^2)$.}
\end{figure}

  Note that $g_{\lambda}(x^2)$, and therefore $g_{\lambda}(x^2)$ has decreased
by a factor of 0.024 when x=4.0 fm.
\vspace{1cm}

\section {Summary and Concluding remarks}

We study the distribution of the nonzero mean squared momentum of quarks
(called the quark virtuality) in the non-perturbative QCD vacuum.
The quark virtuality $\lambda^{2}_{q}$ is given by the
ratio of the local quark-gluon mixed condensate
$<0|:\bar{q}(0)[ig_{s}\sigma G] q(0):|0>$ to the local quark vacuum
condensate $<0|:\bar{q}(0)q(0):|0>$. For our theory of the quark virtuality
distribution, $\lambda^{2}_{q}(x)$, we replace the local $ <0|:\bar{q}D^{2}q:|0>$
by the nonlocal $ <x|:\bar{q}D^{2}q:|0>$. 

 Our calculation shows that as $x$ approaches zero a quark has a finite
nonzero mean squared momentum. In this case, a quark
is similar to a free particle. This indicates asymptotic freedom
of quarks in the QCD vacuum. When $x$ becomes large, say $ x = 10 \rm{fm}$,
$\lambda^{2}_{q}(x)$ approaches zero, which indicates that a quark cannot
move forward. This is a confining phenomena of QCD. Our present prediction 
of the quark mean squared momentum distribution is an initial calculation in 
QCD about properties of the QCD vacuum state. 

 Our  results on quark virtuality at small $x$
and various local vacuum condensates are in good agreement with the 
predictions of other computations such as QCD sum rules, Lattice QCD 
calculations and Instanton model predictions.

\end{document}